\documentclass{article}

\usepackage[final]{neurips_mlsb_2023}

\usepackage[utf8]{inputenc} 
\usepackage[T1]{fontenc}    
\usepackage{hyperref}       
\usepackage{url}            
\usepackage{booktabs}       
\usepackage{amsfonts}       
\usepackage{nicefrac}       
\usepackage{microtype}      
\usepackage{xcolor}         
\usepackage{graphicx}

\title{Enhancing Ligand Pose Sampling for Molecular Docking}

\author{Patricia Suriana, Ron O. Dror \\
Department of Computer Science \\
Stanford University \\
\texttt{psuriana@stanford.edu, rondror@cs.stanford.edu} \\
}

\begin{document}

\maketitle

\begin{abstract}
Deep learning promises to dramatically improve scoring functions for molecular docking, leading to substantial advances in binding pose prediction and virtual screening. To train scoring functions—and to perform molecular docking—one must generate a set of candidate ligand binding poses. Unfortunately, the sampling protocols currently used to generate candidate poses frequently fail to produce any poses close to the correct, experimentally determined pose, unless information about the correct pose is provided. This limits the accuracy of learned scoring functions and molecular docking. Here, we describe two improved protocols for pose sampling: GLOW (auGmented sampLing with sOftened vdW potential) and a novel technique named IVES (IteratiVe Ensemble Sampling). Our benchmarking results demonstrate the effectiveness of our methods in improving the likelihood of sampling accurate poses, especially for binding pockets whose shape changes substantially when different ligands bind. This improvement is observed across both experimentally determined and AlphaFold-generated protein structures. Additionally, we present datasets of candidate ligand poses generated using our methods for each of around 5,000 protein-ligand cross-docking pairs, for training and testing scoring functions. To benefit the research community, we provide these cross-docking datasets and an open-source Python implementation of GLOW and IVES at \url{https://github.com/drorlab/GLOW_IVES}.
\end{abstract}

\section{Introduction}

Protein-ligand molecular docking, which is crucial in drug discovery and molecular modeling \citep{kitchen2004docking,ferreira2015molecular}, predicts the three-dimensional arrangement of ligands within target protein binding sites—a task known as "ligand pose prediction." This computational method is vital for drug candidate exploration and understanding molecular interactions. Conventional docking software relies on sampling algorithms that generate candidate ligand poses based on a given protein structure. This task is inherently difficult due to the multitude of internal conformations the ligand can adopt and the numerous possible ways it can be placed within the protein binding site. Furthermore, a good sampling algorithm must ensure that at least one generated pose closely resembles the experimentally determined "correct pose," which is unknown to the sampling algorithm. Scoring functions then evaluate these poses, selecting candidates predicted to closely match the correct pose. 

While molecular docking has traditionally relied on physics-based scoring functions, recent advances in deep learning, as indicated by studies such as \citep{shen2020machine,francoeur2020three,shen2022boosting,suriana2023flexvdw}, have the potential to revolutionize scoring accuracy. However, the efficacy of deep learning hinges on a crucial factor: generating suitable sets of candidate ligand binding poses. Existing sampling methods often struggle in this regard, frequently failing to produce any correct poses. This challenge intensifies when the protein structure used for docking (the "docking protein structure") significantly differs from the conformation the protein adopts when binding to the query ligand. The inability to sample correct poses creates a twofold problem. Firstly, for an effective deep learning--based scoring function, correct poses must be included in the training dataset to allow models to learn their defining characteristics. However, introducing experimentally determined correct poses, while addressing this need, presents an artificial approach that does not reflect real-world scenarios where such data is unavailable. Moreover, incorporating experimentally determined poses during training could potentially bias the deep learning model's judgment when applied to real-world problems, where all candidate poses, including correct ones, must be generated through sampling. Secondly, the performance of molecular docking relies heavily on the sampling algorithm's ability to consistently yield correct poses. Even with a perfect scoring function, the absence of correct poses among candidates precludes prediction of a correct pose. Hence, there is a pressing need for an enhanced, reliable sampling method capable of consistently generating accurate ligand poses.

To address this challenge, we introduce two improved pose sampling protocols: GLOW (auGmented sampLing with sOftened vdW potential) and a novel method called IVES (IteratiVe Ensemble Sampling). Our protocols substantially increase the likelihood of sampling correct ligand poses, even in scenarios where clashes between the ligand's correct binding pose and the docking protein structure are likely. Importantly, our methods do not rely on information about co-determined ligand poses in the docking protein structure, making them suitable for use with unliganded or predicted protein structures, including those generated by AlphaFold \citep{jumper2021highly, varadi2022alphafold}. Our benchmarking demonstrates that GLOW and IVES effectively enhance ligand pose sampling accuracy for both experimental and AlphaFold-generated protein structures, as measured by the percentage of successful docking cases with correct ligand poses. Additionally, IVES generates multiple protein conformations as part of its workflow, offering considerable value for enhancing geometric deep learning techniques on protein structures and bolstering the robustness of deep learning techniques to small variations around correct poses in the context of protein-ligand docking.

To encourage broader engagement and utilization within the research community, we have created carefully curated datasets containing candidate ligand poses generated using our improved sampling methods. These datasets comprise approximately 5,000 protein-ligand cross-docking pairs, serving as invaluable resources for training and evaluating scoring functions. To promote widespread access and utilization, we have made available an open-source Python implementation of GLOW and IVES, along with the newly developed cross-docking datasets. These resources can be accessed at \url{https://github.com/drorlab/GLOW_IVES}.

\section{Related Works}
\begin{figure}[!htbp]
\centering
\includegraphics[width=0.6\textwidth]{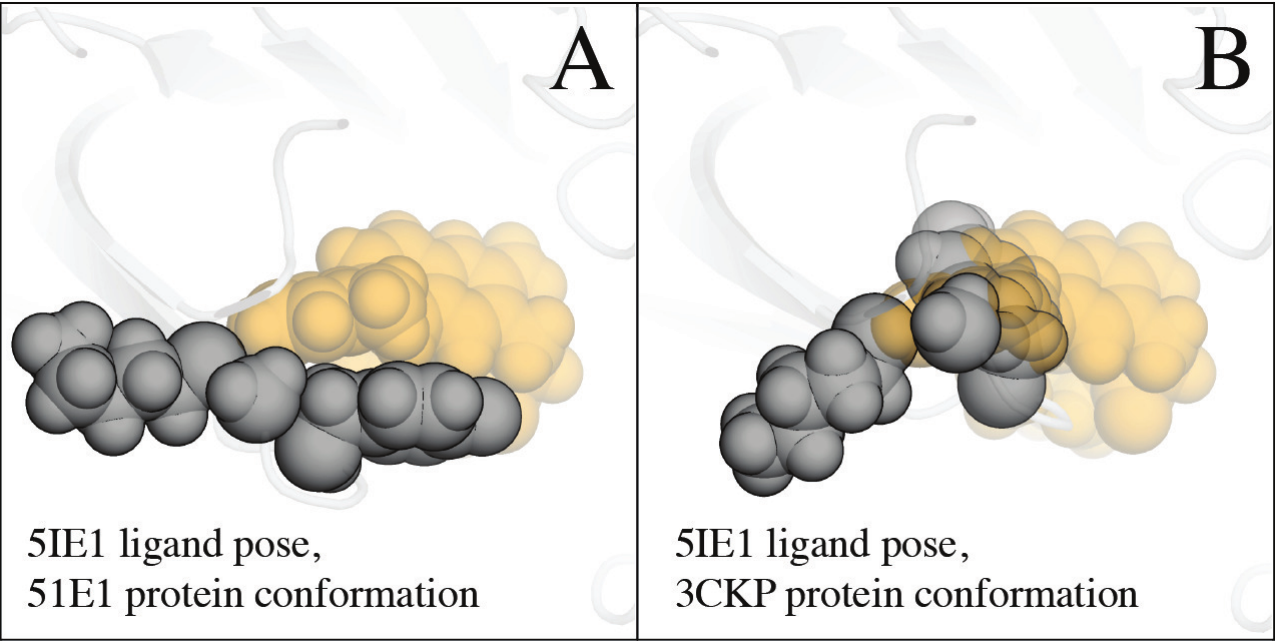}
\caption{The native binding pose of a ligand often clashes with the experimentally determined structure of its target protein, especially when that structure features a different ligand. In Panel A, we observe the structure of $\beta$-secretase (BACE-1), a key drug target, bound to "compound 5" (PDB entry 5IE1 \citep{jordan2016fragment}). Here, the ligand (depicted as orange spheres, each representing an atom) packs favorably against two protein amino acids (gray spheres) in the binding pocket with no clashes. In contrast, Panel B presents the same ligand (compound 5) in an identical geometry, but overlaid on a BACE-1 structure determined in the presence of a different ligand (PDB entry 3CKP, \citep{park2008synthesis}). In this case, the two amino acids (gray spheres) adopt different positions, resulting in significant clashes with the ligand atoms (orange spheres).}
\label{fig:native_holo}
\end{figure}

Numerous deep learning techniques have emerged to score candidate poses in molecular docking, traditionally relying on datasets generated through rigid protein docking, which assume fixed protein structures during sampling \citep{verdonk2003improved, friesner2004glide, allen2015dock, forli2016computational}. For example, the CrossDock2020 model \citep{francoeur2020three} is based on poses generated by Smina's rigid protein docking. However, this method falls short when adjustments are needed in the docking protein structure to accommodate the correct ligand binding pose, resulting in clashes and rejection of the correct pose during rigid docking (see Figure \ref{fig:native_holo}). This limitation in sampling correct poses presents a significant challenge and constrains the performance of scoring functions, including those based on deep learning.

To address this, flexible protein docking methods consider protein flexibility during sampling \citep{jones1997development,lemmon2012rosetta,sherman2006novel,miller2021reliable}. Strategies include alternating ligand and protein sampling steps or temporarily substituting flexible residues with alanine. Some, like Schrödinger IFD-MD \citep{miller2021reliable}, enhance accuracy by incorporating experimentally co-determined ligand poses. Conversely, certain recent deep-learning approaches, such as \citet{krishna2023generalized}, directly generate the protein-ligand complex. However, substantial computational resources are essential for all these methods, and flexible docking methods place an added emphasis on accurately selecting flexible residues. In practical scenarios, both flexible protein docking and deep-learning-based approaches consistently demonstrate lower accuracy compared to rigid protein docking \citep{ravindranath2015autodockfr,bender2021practical,krishna2023generalized}.

Flexible protein docking methods and deep-learning-based approaches have been explored to enhance ligand binding pose sampling, but they bring their own limitations, including computational costs and reliance on experimental data. The introduction of GLOW and IVES seeks to tackle these challenges, providing a promising path to improve ligand pose sampling accuracy and efficiency in protein-ligand docking, which could benefit the development and evaluation of deep learning-based scoring functions.

\section{Methods}

\subsection{Improved ligand pose sampling protocols for docking}

A significant drawback of rigid protein docking is its inability to generate correct pose when it clashes with the docking protein structures. The ligand's correct pose receives a poor score due to these clashes, leading to exceptionally high calculated van der Waals (VDW) energy values. To this end, we introduce GLOW. GLOW enhances rigid protein docking by incorporating poses generated with a softened VDW potential alongside those using a normal VDW potential.

\begin{figure}[!htbp]
\centering
\includegraphics[width=\textwidth]{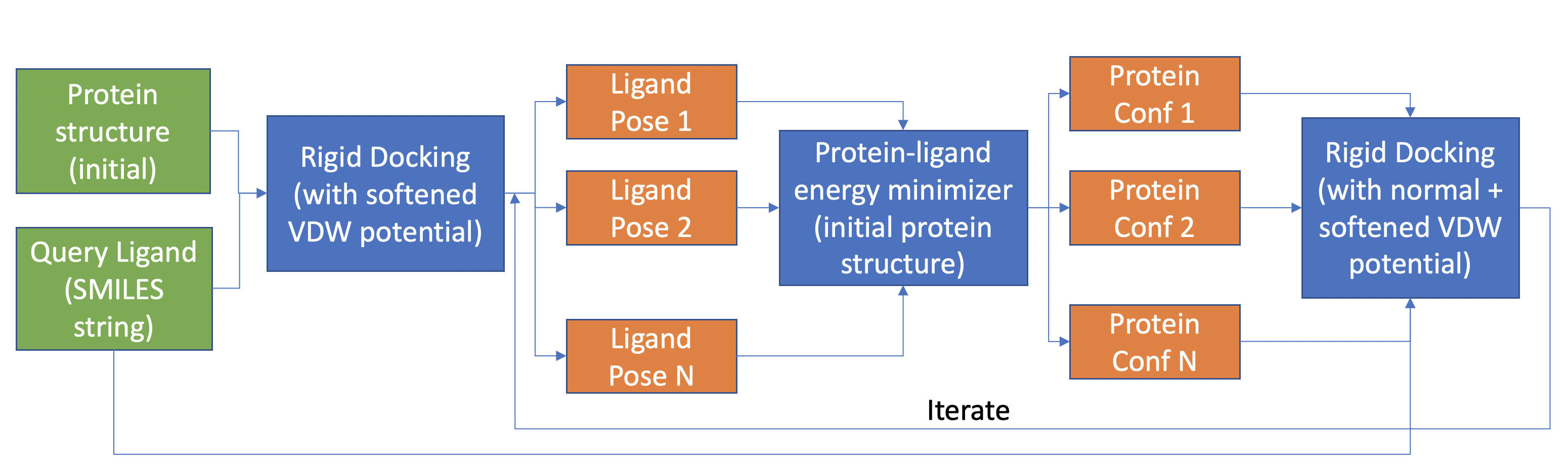}
\caption{Schematic of the IVES workflow. Initially, we perform rigid protein docking using softened VDW potentials, yielding initial ligand poses. Due to this softening, some poses may clash with the docking structure. Then, we select the top N poses as "seed poses" for guiding the minimization of the input docking structure, producing an ensemble of N protein conformations. Residues within 8\AA\ of the ligand pose are allowed to move, while the rest remain fixed, including the ligand. Parallel rigid docking with normal and softened VDW potential of the input query ligand onto these N conformations follows. This process may be iterated, but typically one iteration suffices, as further iterations offer minimal improvements.}
\label{fig:ives}
\end{figure}

Furthermore, we present IVES, an innovative approach to enhance ligand pose sampling accuracy in protein-ligand docking (see Figure \ref{fig:ives}). IVES incorporates a combination of alternating protein-ligand pose sampling strategies inspired by flexible protein docking, all while utilizing both normal and softened VDW potentials. IVES begins with rigid protein docking, using a softened VDW potential to create initial ligand poses, allowing some clashes with the docking structure. The "seed poses," selected from the top N poses in this initial set based on docking scores or alternative scoring functions for assessing protein-ligand docked poses, guide the minimization of the input docking structure within an 8\AA\ radius of the ligand pose, while keeping the ligand pose and other residues fixed. Subsequently, the input ligand is redocked onto these N protein conformations, employing both normal and softened VDW potentials independently for each conformation, allowing for parallelization to accelerate the process. If necessary, this step can iterate by merging poses from the prior iteration and selecting the top N ligand poses for the next iteration, although a single iteration often suffices as subsequent ones provide marginal improvements in practice.

To make our approaches accessible to a broader scientific community, we built GLOW and IVES on top of Smina \citep{koes2013lessons} and OpenMM \citep{eastman2017openmm}, open-source software for molecular docking and protein structure minimization, respectively. In general, our approaches can be built on top of any existing software for rigid protein docking or protein minimization. 

\subsection{Datasets} \label{sec:datasets}

We offer datasets of candidate ligand poses generated by GLOW and IVES, potentially valuable for training machine learning-based docking. Our dataset includes around 4,102 protein-ligand cross-docking pairs derived from 238 unique proteins, utilizing protein structures from the high-resolution PDBBind 2019 refined dataset \citep{wang2005pdbbind,liu2015pdb}. Additionally, we provide testing datasets categorized into four groups: (1) "typical (experimental)" with 322 experimentally determined pairs (from \cite{paggi2021leveraging}), (2) "challenging (experimental)" with 258 pairs requiring significant docking protein structure adjustments to fit the ligand (from \cite{miller2021reliable}), (3) "typical (AlphaFold)" features the same pairs as "typical (experimental)" but uses AlphaFold 2 protein models for docking (totaling 322 pairs), and (4) "challenging (AlphaFold)" features the same pairs as "challenging (experimental)" but uses AlphaFold 2 protein models instead of experimentally determined structures for docking (totaling 179 pairs). For more details, refer to \ref{sup:datasets}.

\section{Results}

\subsection{Evaluation of the sampling performance on the test sets} \label{sec:test}
We evaluated the sampling performance of GLOW and IVES on test sets by measuring the percentage of cross-docking cases that yielded any correct pose. A correct pose was defined as having a root mean square deviation (RMSD) from the experimentally determined pose equal to or less than 2.0 \AA, a widely accepted practical threshold \citep{kontoyianni2004evaluation,cole2005comparing}. For reference, we included two baseline methods: (1) "Default," representing typical docking scenarios, generating a maximum of 20 poses \citep{francoeur2020three}; (2) "Default, max poses," allowing the maximum number of poses, representing the upper limit of the docking protocol. For GLOW, we enabled the generation of as many poses as possible. IVES produced poses using 20 protein conformations in a single iteration and generated a maximum of 300 poses for each protein conformation. Both GLOW and IVES are implemented on Smina. For consistency, we used Smina for the baseline methods as well. Additional settings details can be found in \ref{sup:docking_settings}.

\begin{figure}[!h]
\centering
\includegraphics[width=0.95\textwidth]{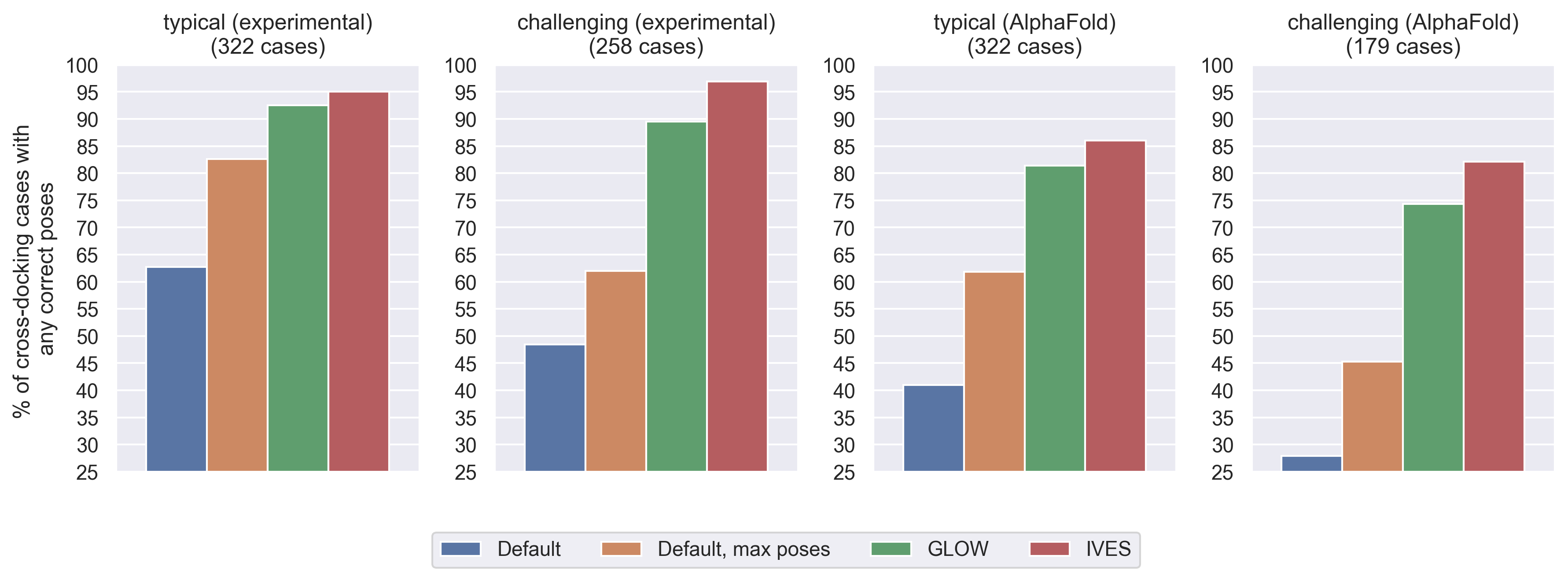}
\caption{Sampling performance of GLOW and IVES on test sets, measured by the percentage of cross-docking cases with any correct pose. GLOW (green) and IVES (red) consistently outperform the baseline methods "Default" (blue) and "Default, max poses" (orange), especially for the challenging and AlphaFold benchmarks where the protein structure undergoes substantial conformational changes upon binding to the ligand, differing from the structure employed during the docking process.}
\label{fig:sampling_perf}
\end{figure}

Overall, GLOW and IVES consistently outperformed baseline methods, especially in challenging and AlphaFold benchmarks where the protein structure undergoes significant conformational changes upon binding to the ligand, differing from the structure employed for docking (see Figure \ref{fig:sampling_perf}). These results highlight their potential to enhance the accuracy of pose sampling in protein-ligand docking applications.

\subsection{Comparing IVES and GLOW to flexible protein docking}

We compared the performance of GLOW and IVES with the open-source Smina flexible protein docking ("Smina flexible"), specifically focusing on cross-docking cases where "Smina flexible" completed within 48 hours on a single CPU, constituting 40\% of the dataset (see Figure \ref{supfig:percentage_of_any_correct-flex_dock_include_failures_12000} for details). Despite this selective assessment, GLOW and IVES consistently outperformed "Smina flexible," especially in challenging and AlphaFold benchmarks (Figure \ref{supfig:percentage_of_any_correct-flex_dock_12000}). Using 20 protein conformations, GLOW and IVES demonstrated significantly faster runtimes---approximately 20 minutes and 6-7 hours, respectively---on a single CPU compared to "Smina flexible," which typically required 16 hours. While IVES's runtime scales linearly with the number of protein conformations in a serialized setting, it exhibits high parallelizability, completing in around 20 minutes on average when fully parallelized. Moreover, IVES offers extensive customization, allowing users to adjust the sampling process by specifying the number of protein conformations and the maximum number of generated poses per docking run for each conformation. This flexibility enables a balance between thoroughness and computational costs.

IVES also achieved comparable sampling performance to Schrödinger IFD-MD, a proprietary state-of-the-art flexible protein docking software, using only 20 protein conformations versus IFD-MD's 1000 conformations (Figure \ref{supfig:schrodinger_ifd_md_vs_ives}). Notably, IVES---unlike IFD-MD---does not rely on an experimentally co-determined ligand pose in the docking protein structure, which allows IVES to work with unliganded or predicted structures such as those from AlphaFold, broadening its applicability.

Overall, our results not only highlight the competitive sampling capabilities of GLOW and IVES, which in some cases outperformed flexible protein docking, but also underscores their value in scenarios where access to experimentally determined ligand poses within the docking protein structure is unavailable.

\section{Discussion}

Our benchmark results demonstrate the substantial improvements achieved by GLOW and IVES in increasing the probability of sampling correct poses in protein-ligand docking. These gains are especially notable in challenging and AlphaFold benchmarks, where protein structures exhibit significant conformational differences when bound to query ligands compared to those used in docking. Additionally, IVES generates multiple protein conformations, which can be beneficial for geometric deep learning on protein structures. Furthermore, we provide datasets of candidate ligand poses generated by our methods for approximately 5,000 protein-ligand cross-docking pairs. These datasets may serve as valuable resources for developing and assessing deep-learning-based scoring functions in molecular docking.

While IVES demonstrates the best performance, its sampling efficiency is contingent upon the quality of the initial seed poses. Ideally, these initial poses should closely resemble experimentally determined poses to reduce the need for generating numerous protein conformations during sampling, as illustrated in Figure \ref{supfig:num_prot_conf_vs_percent_hit}. The selection of these poses relies on an effective scoring function, creating a complex interplay between scoring and sampling. Nevertheless, IVES offers a customizable workflow for ligand pose sampling, allowing the generation of improved samples to train a machine-learned protein-ligand docking scoring function. This scoring function, in turn, can refine seed pose selection in IVES, establishing a dynamic feedback loop that continuously improves pose sampling and scoring accuracy.

\section*{Acknowledgment}
Thanks to Joseph M. Paggi for helpful discussions.

\section*{Funding Information}
This work was supported by National Institutes of Health (NIH) grant R01GM127359 (R.O.D.) and a US National Science Foundation (NSF) Graduate Research Fellowship (P.S.).

\bibliographystyle{plainnat}
\bibliography{neurips_2023}

\setcounter{section}{0}
\renewcommand{\thesection}{S\arabic{section}}
\renewcommand{\theHsection}{Supplement.\thesection}

\setcounter{figure}{0}
\renewcommand\thefigure{S\arabic{figure}}
\renewcommand{\theHfigure}{Supplement.\thefigure}

\clearpage
{\Huge{Supplementary Information}}

\section{Datasets} \label{sup:datasets}
To create ligand pose datasets from the PDBBind 2019 refined dataset \citep{wang2005pdbbind,liu2015pdb}, we initially categorized protein-ligand complex structures based on the protein, grouping structures of the same target protein. To ensure a balanced representation in our dataset and prevent bias from overrepresented proteins with numerous experimentally determined structures bound to various ligands, we randomly selected a maximum of 15 structures for each protein to serve as docking protein structures. Furthermore, the selection of protein-ligand cross-docking pairs took into account the Tanimoto similarity between the query ligand to be docked and the co-determined ligand present in the docking structure \citep{bajusz2015tanimoto}, with a threshold of less than 0.4, indicating dissimilarity. For each unique ligand, we chose up to 5 protein structures as cross-docking pairs. This resulted in 4,102 distinct protein-ligand cross-docking pairs (derived from 238 unique proteins), ensuring diversity and relevance in our dataset.

As outlined in \ref{sec:datasets}, the "challenging (AlphaFold)" dataset includes the same cross-docking pairs as "challenging (experimental)," but it employs AlphaFold protein structures for docking, in contrast to experimentally determined ones. However, the "challenging (experimental)" dataset contains cross-docking pairs with structures from the same protein but bound to different ligands, often referred to as "holo structures." Consequently, the "challenging (AlphaFold)" dataset has fewer cross-docking pairs available when using AlphaFold structures for docking, as those holo structures of the same protein will be mapped to the same AlphaFold structure.

All the datasets contain poses generated through GLOW and IVES. We followed the protocols described in \citet{paggi2021leveraging} for preparing protein-ligand complex structures and ligands prior to the docking process. IVES produced poses using 5 protein conformations in a single iteration, generating a maximum of 300 poses for each docking to each protein conformation. Seed poses are selected based on the RTMscore \citep{shen2022boosting}.

\section{Docking settings} \label{sup:docking_settings}

In Figure \ref{fig:sampling_perf}, all pose sampling methods were run with the following settings: an exhaustiveness value of 16, a minimum RMSD filter set at 1.5\AA\, and a search space box sized at 20\AA\, with its center aligned to the bound ligand pose within the docking protein structure. In cases involving AlphaFold-generated structures, the search space center was determined using an experimentally determined ligand pose (not the query ligand). In practical scenarios where experimentally resolved ligand poses are unavailable, search space determination can be facilitated using binding pocket finder tools. Additionally, to create softened VDW potentials for GLOW and IVES, we adjusted the repulsion weight to 0.2, departing from the default value of 0.840.

\newpage
\section{Distributions of the number of sampled poses across different methods} \label{sup:pose_dist}

\begin{figure}[!htb]
\centering
\includegraphics[width=1\textwidth]{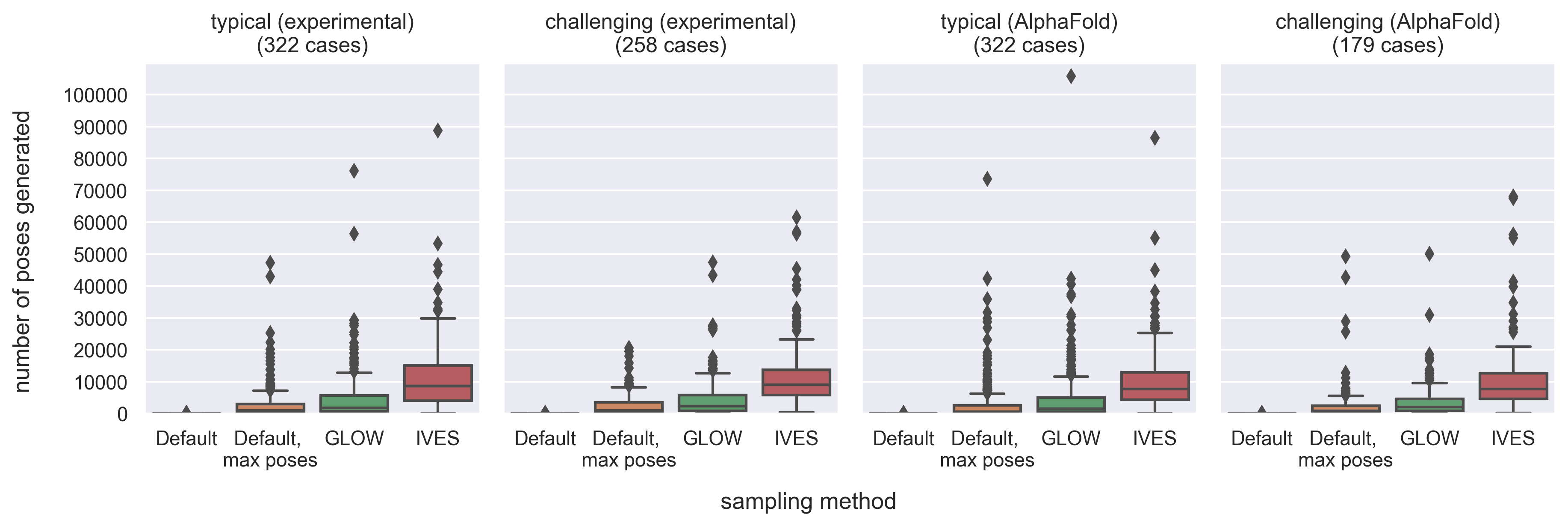}
\caption[Distribution of the number of poses generated by GLOW and IVES compared to baseline methods "Default" and "Default, max poses"]{Distribution of the number of poses generated by GLOW and IVES compared to baseline methods "Default" and "Default, max poses". To ensure a fair comparison between IVES, GLOW, and "Default, max poses," we allowed "Default, max poses" and "GLOW" to generate as many poses as possible (specifically, up to 1 million poses per ligand for "Default, max poses" and up to an additional 1 million poses per ligand with the softened VDW potential in GLOW). Nevertheless, IVES typically generated more poses than these other methods, because IVES utilizes multiple protein conformations, expanding the feasible pose landscape.
}
\label{supfig:num_poses_dist-normal_soft_seed_20}
\end{figure}

\section{Comparison of GLOW and IVES sampling performances with Smina flexible protein docking} \label{sup:smina_flex}

\begin{figure}[!htb]
\centering
\includegraphics[width=1\textwidth]{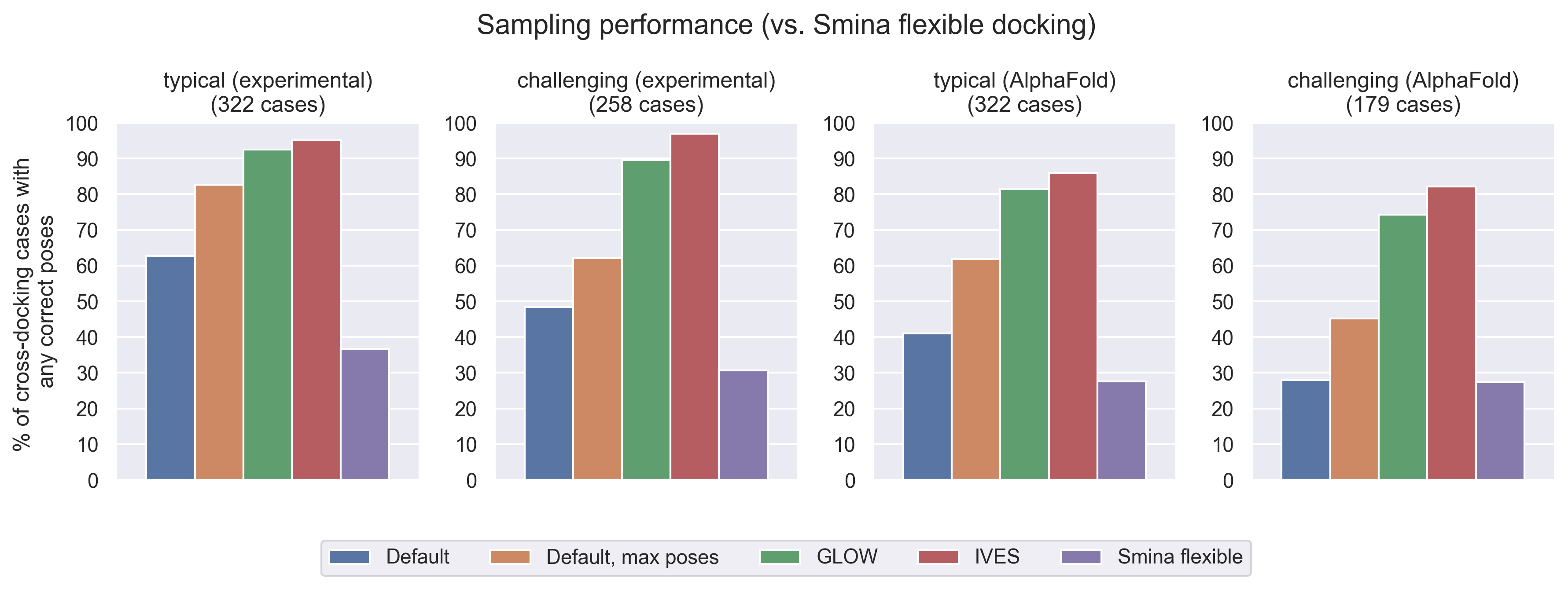}
\caption[Sampling performance of GLOW and IVES compared to Smina flexible protein docking on the test sets]{Sampling performance of GLOW and IVES compared to Smina flexible protein docking on the test sets, measured by the percentage of cross-docking cases with at least one correct pose. In this analysis, we compare the performance of the GLOW and IVES, as described in Figure \ref{fig:sampling_perf}, with Smina flexible protein docking ("Smina flexible"). Similar to Figure \ref{fig:sampling_perf}, "Smina flexible" is run with a search space box of size 20\AA\ centered around the bound ligand pose in the protein structure used for docking. We aim to achieve a similar pose count for "Smina flexible" whenever possible. It's important to note that "Smina flexible" operates under a 48-hour time limit, with runs exceeding this duration considered failures. Among the results, 20\% of "Smina flexible" runs did not complete within 48 hours, 40\% completed within the timeframe but failed to generate poses, while the remaining 40\% completed within 48 hours and successfully generated poses. These statistics collectively contribute to the relatively inferior performance of "Smina flexible" compared to other methods, including the baseline approaches "Default" and "Default, max poses."}
\label{supfig:percentage_of_any_correct-flex_dock_include_failures_12000}
\end{figure}

\newpage

\begin{figure}[!htb]
\centering
\includegraphics[width=1\textwidth]{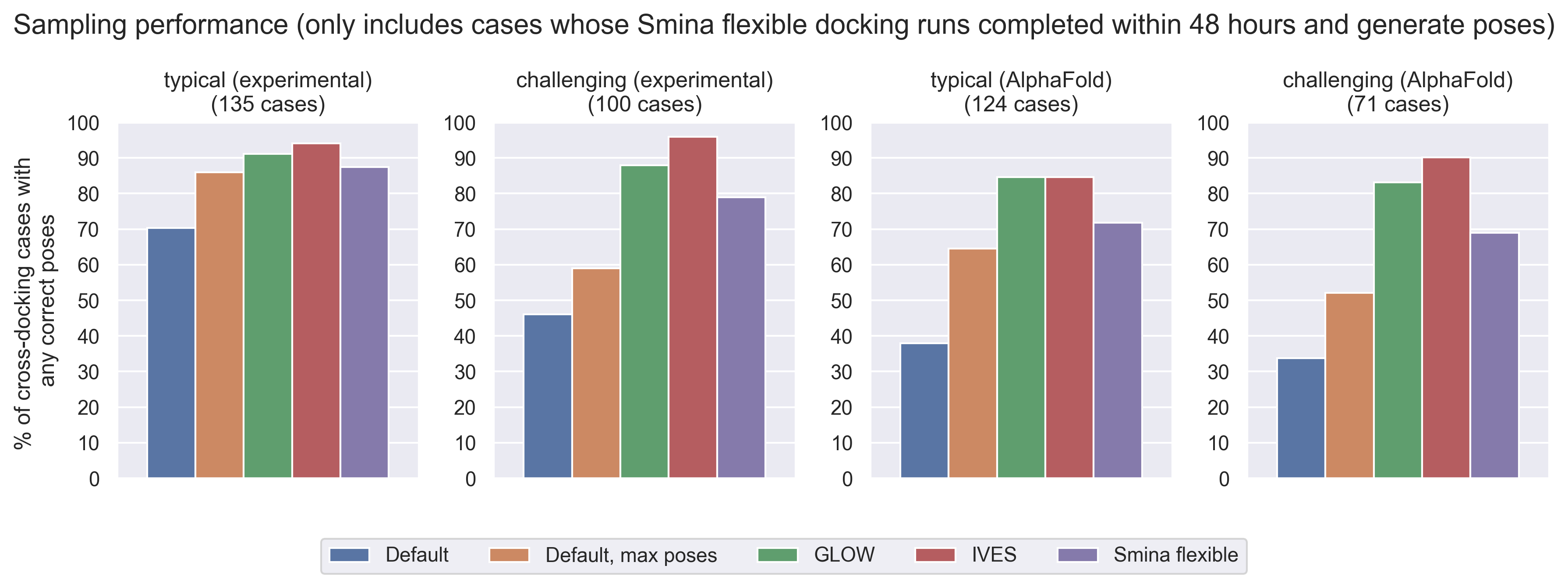}
\caption[Sampling performance of GLOW and IVES compared to Smina flexible protein docking across multiple datasets on cross-docking cases where "Smina flexible" successfully completed within 48 hours on one CPU and generated poses]{Sampling performance of GLOW and IVES compared to Smina flexible protein docking across multiple datasets, measured by the percentage of cross-docking cases with at least one correct pose, focusing on cases where "Smina flexible" completed within 48 hours on one CPU and generated poses. This analysis differs from that of Figure \ref{supfig:percentage_of_any_correct-flex_dock_12000} in that we only consider cross-docking cases where "Smina flexible" successfully completed within 48 hours on one CPU and generated poses, accounting for approximately 40\% of the total cases. Even in this subset, both GLOW and IVES consistently outperform "Smina flexible", particularly in challenging and AlphaFold benchmarks where the protein structure undergoes substantial conformational changes upon binding to the ligand, differing from the structure employed during the docking process. In addition, GLOW and IVES are considerably faster than Smina flexible protein docking. On average, GLOW finishes in about 20 minutes, while IVES typically takes 6-7 hours on a single CPU. In contrast, "Smina flexible" runs, completed within a 48-hour timeframe, average around 16 hours. It's worth highlighting IVES' high parallelizability, achieving an average completion time of approximately 20 minutes when fully parallelized. Furthermore, IVES offers extensive customization options, allowing users to adjust sampling thoroughness by selecting the number of protein conformations or setting the maximum number of generated poses per docking with each conformation. This flexibility empowers users to strike a balance between thoroughness and computational costs.}
\label{supfig:percentage_of_any_correct-flex_dock_12000}
\end{figure}

\begin{figure}[!htb]
\centering
\includegraphics[width=1\textwidth]{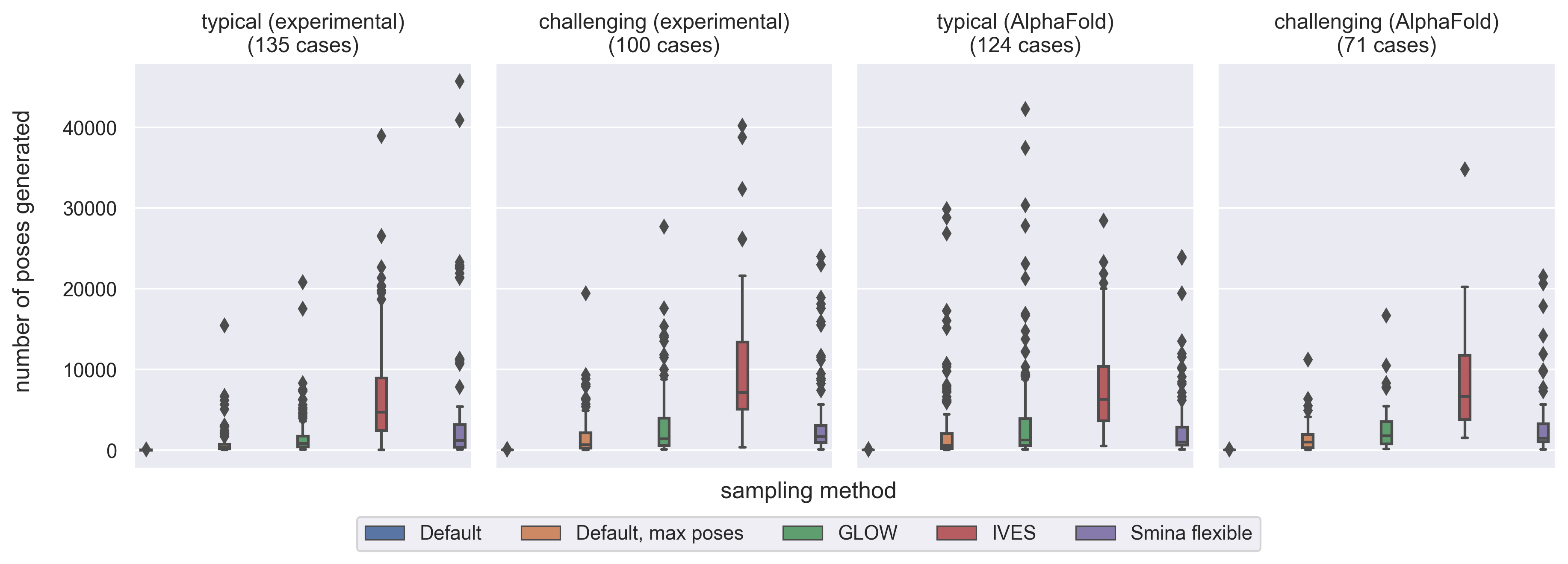}
\caption[Distribution of the number of poses generated by GLOW and IVES compared to Smina flexible docking ("Smina flexible") and baseline methods "Default" and "Default, max poses"]{Distribution of the number of poses generated by GLOW and IVES compared to Smina flexible docking ("Smina flexible") and baseline methods "Default" and "Default, max poses". Here, we only consider cross-docking cases where "Smina flexible" successfully completed within 48 hours on one CPU and generated poses, accounting for approximately 40\% of the total cases. To ensure a fair comparison, we allowed "Default, max poses" and "GLOW" to generate as many poses as possible. }
\label{supfig:num_poses_dist-flex_dock_12000}
\end{figure}

\newpage

\section{Comparison of IVES sampling performance with Schrödinger IFD-MD flexible protein docking} \label{sup:ifd_md}

\begin{figure}[!htb]
\centering
\includegraphics[width=0.4\textwidth]{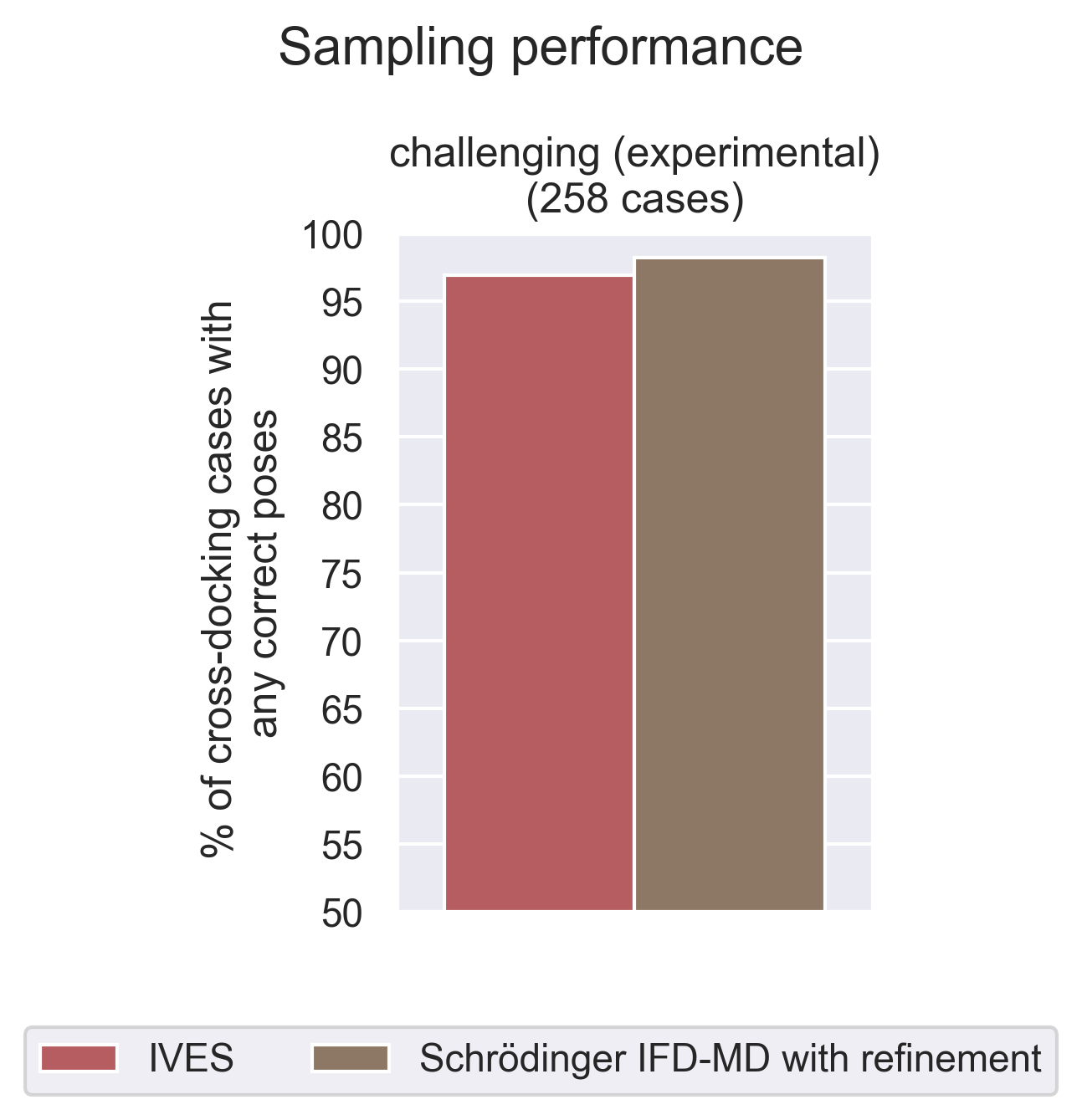}
\caption[Sampling performance of IVES vs. Schrödinger IFD-MD on "challenging (experimental)" dataset]{Sampling performance of IVES vs. Schrödinger IFD-MD on "challenging (experimental)" dataset. Here we compare the sampling performance (measured by the percentage of cross-docking cases with at least one correct pose) of IVES versus Schrödinger IFD-MD (with refinement). IVES (in red) exhibited comparable performance to Schrödinger IFD-MD (in brown), despite using only 20 protein conformations compared to IFD-MD's 1000. Notably, IVES doesn't rely on an experimentally co-determined ligand pose bound in the docking structure, making it applicable to docking to unliganded or predicted structures such as those generated by AlphaFold.}
\label{supfig:schrodinger_ifd_md_vs_ives}
\end{figure}

\newpage

\section{The impact of the number of protein conformations and seed pose quality on IVES sampling performance} \label{sup:efficiency}

\begin{figure}[!htb]
\centering
\includegraphics[width=1\textwidth]{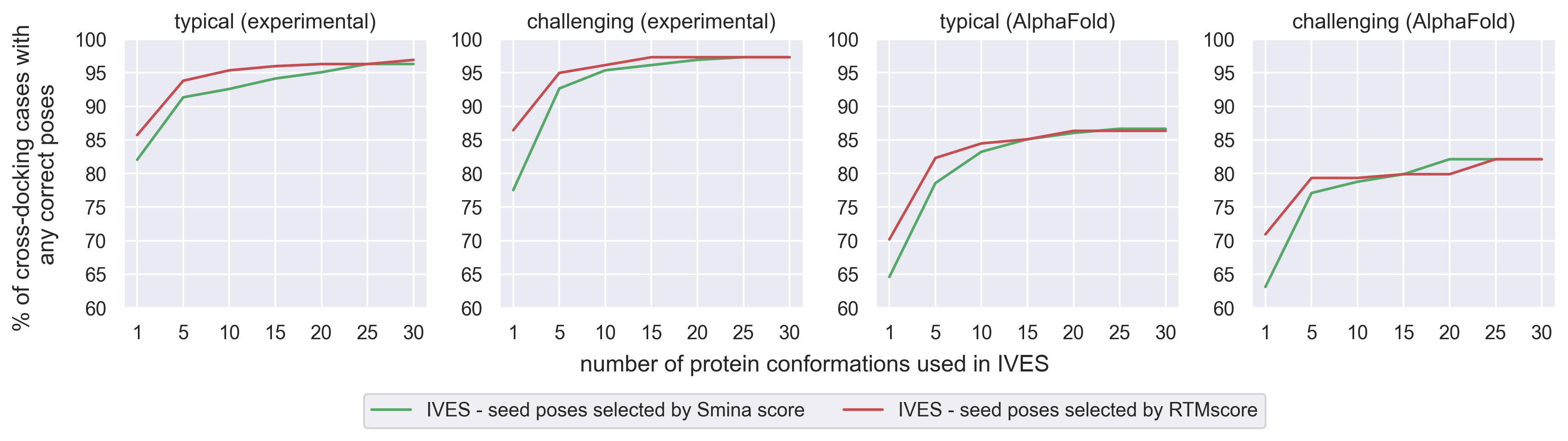}
\caption[Sampling performance of IVES as a function of the number of protein conformations used for sampling]{Sampling performance of IVES as measured by the percentage of cross-docking cases with at least one correct pose as a function of the number of protein conformations used for sampling. As we increase the number of protein conformations employed by IVES, we observe a significant increase in the percentage of cross-docking cases yielding correct poses. This increase is most noticeable when using 1 to 5 protein conformations. It's important to note that IVES runtime scales proportionally with the number of protein conformations, but its high parallelizability efficiently utilizes computational resources. Therefore, for those with computational constraints, running IVES with 5 protein conformations strikes a favorable balance between resources and sampling performance.
Efficient IVES sampling relies on the quality of seed poses chosen for generating protein conformations. Better seed poses, ideally close to the "correct" pose, reduce the need for large number of protein conformations, thus lowering computational demands. We employ both the Smina docking score and RTMscore \citep{shen2022boosting}, a machine-learned scoring function for ranking ligand poses, to rank and select seed poses. In our evaluation, RTMscore emerges as the better choice for ranking, enhancing IVES's sampling efficiency with fewer protein conformations compared to when using Smina docking score (highlighted in red vs. green). This emphasizes the critical role of seed pose quality in optimizing IVES's sampling outcomes.
}
\label{supfig:num_prot_conf_vs_percent_hit}
\end{figure}



\end{document}